\documentclass{elsart}
\usepackage{amssymb}
\usepackage{graphics}
\journal{Solid State Communications}
\begin{document}
\begin{frontmatter}
\title{Locality properties and Wannier functions for interacting systems}
\author{Erik Koch}
\ead{E.Koch@fkf.mpg.de}
\address{Max-Planck Institut f\"ur Festk\"orperforschung,
         Heisenbergstra\ss e 1, 70569 Stuttgart, Germany}
\author{Stefan Goedecker}
\ead{goedecke@drfmc.ceng.cea.fr}
\address{D\'epartement de recherche fondamentale sur la mati\`ere
           condens\'ee, SP2M/NM, CEA-Grenoble, 38054 Grenoble cedex~9, France}
\date{\today}
\begin{abstract}
We define Wannier functions for interacting systems, and show that the results 
on the localization of the Wannier functions for non-interacting systems carry 
over to the Wannier functions for interacting systems. In addition we 
demonstrate that the characterization of metals and insulators by the decay
properties of their respective density matrices does not only apply to
non-interacting, but also to interacting systems.
As a prototypical example of a correlated system we investigate the 
one-dimensional Hubbard model. We propose an expression for the density 
matrix of that model, and derive a relation between the decay constant of 
the density matrix and the gap.
\end{abstract}
% 71.10.-w Theories and models of many electron systems
% 71.30.+h Metal-insulator transitions and other electronic transitions
% 71.10.Fd Lattice fermion models (Hubbard model, etc.)
\end{frontmatter}

The locality properties of solids, their so called nearsightedness, 
have recently moved to the focus of much attention 
\cite{Kohn96,Arias,Baer,Marzari,GoedeckerRMP,Ordejon,Resta}.
This is largely due to the intense efforts going into the development of
electronic structure methods that scale linearly with system size 
\cite{GoedeckerRMP,Ordejon}. These methods depend crucially on the locality 
of the density matrix. The closely related issue of the locality of the Wannier
functions \cite{Blount} has also attracted renewed interest, largely due
to the development of methods for the practical construction of localized
Wannier functions \cite{Marzari}. 
The uses of Wannier functions are well proven. First,
they are fundamental in the theory of electron dynamics in the
presence of weak external fields \cite{NenciuRMP}.
Second, they allow for an intuitive interpretation of the bonding properties
in solids \cite{Marzari}: Localized Wannier functions correspond to either
bonds or lone electron pairs. 
Third, they are at the center of the modern theory of polarization 
\cite{KingSmith}.
Finally, they are important for some linear scaling algorithms
\cite{GoedeckerRMP,Ordejon}.

General results on the localization of the Wannier functions are quite 
difficult to obtain. The problem has actually been called ``one of the 
few basic questions of the quantum theory of periodic solids in the 
one-electron approximation which is not completely solved'' \cite{Nenciu}.
The most important results so far are the proofs of the existence of
exponentially localized Wannier functions for isolated, simple bands
in any dimension \cite{Nenciu}, and for complex bands in the 
tight-binding limit, and in perturbation theory \cite{desCloizeauxII}.
For practical applications knowledge of the decay constants is
of considerable importance. First results were obtained in \cite{Kohn59}, 
considerably more general results for the density matrix were given 
in \cite{Arias}. 

Trying to extend the results to interacting systems we are faced with
a fundamental problem: The Hamiltonian is no longer a one-body operator.
Hence it seems to be impossible to define meaningful single-electron Bloch 
waves and the corresponding Wannier functions. This is true already in the 
limit of weak interactions, although there one would expect that the 
non-interacting Wannier functions still could be useful. We can, however, 
avoid the problem of having to
deal with a many-body operator, by changing perspective. Looking at the
one-body density matrix, we realize that it can replace the 
(single-electron) Hamiltonian in the standard construction of Wannier 
functions. We are thus led to Wannier functions made from 
natural orbitals \cite{NatOrbt}, which we christen natural Wannier functions.
Instead of the energy bands $\varepsilon_j({\bf k})$
we now consider occupation bands $n_j({\bf k})$. We prove that 
the projection operator $P({\bf k})$ onto an isolated set of 
occupation bands is analytic in ${\bf k}$. 
Once this result is established, 
the results \cite{Nenciu,desCloizeauxII} obtained for the standard Wannier 
functions immediately carry over to the natural Wannier functions. 
So the natural Wannier functions can be considered as the natural
generalization of the concept of Wannier functions to interacting systems.
Being constructed from natural orbitals, they are an optimal basis, meaning 
that it is sufficient to consider only the bands $n_j({\bf k})$ with high 
occupation to obtain a good description of the interacting system 
\cite{NatOrbt}.
Being Wannier functions, they are localized, thus allowing to take advantage
of $O(N)$ methods in evaluating, e.g., Coulomb matrix elements. Moreover, 
they can be expected to represent the chemical bonding in the correlated 
system, as do the ordinary Wannier functions in the independent-particle case.
We furthermore show that the characterization of metals and
insulators in terms of the decay properties of the density-matrix 
\cite{desCloizeauxI}, also applies to correlated systems.
Thus the decay constant $\gamma$ may 
play a similar role as the localization length defined in \cite{Resta}. 
Finally, we consider the one-dimensional Hubbard model as an explicit 
example of a correlated system, deriving a relation giving the decay
constant $\gamma$ as a function of the gap, thus extending the results
of \cite{Arias} to an interacting system.

We start our construction from the many-body wavefunction
$\Psi  ({\bf r}_1, {\bf r}_2, ... ,{\bf r}_N)$.
For simplicity we consider spinless electrons. 
The one-body density matrix $D$ is then given by
\begin{displaymath}
D({\bf r},{\bf r}') =\int d{\bf r}_2 \cdots d{\bf r}_N 
    \Psi^\ast ({\bf r}', {\bf r}_2, \ldots {\bf r}_N) 
    \Psi      ({\bf r} , {\bf r}_2, \ldots {\bf r}_N) .
\end{displaymath}
Its eigenfunctions are the natural orbitals, its eigenvalues the natural 
occupation numbers. 
Clearly, $D$ is Hermitian. Furthermore, translating all the spatial arguments
of $\Psi$ by a lattice vector ${\bf R}$ multiplies the wavefunction by a phase 
factor. Since these phase factors cancel inside the density matrix, we find
$D({\bf r}+{\bf R},{\bf r}'+{\bf R}) = D({\bf r},{\bf r}')$.
Thus the Bloch theorem applies and, in analogy to the Hamiltonian of a 
periodic solid in an independent-particle picture \cite{AshcroftMermin}, 
the eigenvalues of the density matrix form bands $n_j({\bf k})$, the 
occupation bands, with the corresponding natural orbitals being Bloch functions
\begin{equation}\label{natk}
\Phi_{j,{\bf k}}({\bf r}) = e^{i{\bf kr}}\:U_{j,{\bf k}}({\bf r})\:,
\end{equation}
where $j$ is the band index, ${\bf k}$ a vector in the Brillouin 
zone, and the $U_{j,{\bf k}}({\bf r})$ are periodic functions with respect 
to the real space primitive cell. They are the eigenvectors of the
${\bf k}$-dependent density matrix
\begin{displaymath}
D_{\bf k}=\sum_j n_j({\bf k})\: |U_{j,{\bf k}}\rangle\langle U_{j,{\bf k}}|\:.
\end{displaymath}
Using the standard prescription \cite{AshcroftMermin}, we can then
construct Wannier functions, which, for obvious reasons, we call 
natural Wannier functions:
\begin{equation}
 W_j({\bf r-R}) = {V\over(2\pi)^3}
   \int_{BZ} d{\bf k}\,e^{i{\bf k}({\bf r-R})}\,U_{j,{\bf k}}({\bf r})\:;
\end{equation}
here $V$ is the volume of the real-space cell and the integration is over
the Brillouin zone (BZ).

The construction of the natural Wannier functions is very similar to that
of the conventional Wannier functions for a non-interacting system, the
density matrix $D$ taking the place of the non-interacting Hamiltonian.
In the limit of vanishing interaction, the natural Wannier functions do,
however, not reduce to the conventional Wannier functions. This is clear,
since in the limit of no interaction the density matrix becomes the projector
onto the occupied subspace; i.e.\ all occupied (unoccupied) states are 
degenerate with eigenvalue one (zero). By degenerate perturbation theory,
in the limit of vanishing interaction, the 
natural Wannier functions will therefore have to diagonalize the first
term in the perturbation expansion in their respective subspace (occupied or
unoccupied) \cite{NatOrbt}. They are therefore generalized Wannier functions 
in the sense of \cite{Marzari}, which, respectively, span the space of occupied 
and unoccupied states. 
%% could be removed
 Clearly, using several bands increases the flexibility to construct more  
 localized Wannier functions, since one can take advantage of the unitary
 transformations allowed in the space spanned by the Bloch functions of
 different bands. In the extreme limit, where we allow all bands, it is
 obviously possible to construct Wannier functions that are perfectly
 localized delta functions.
%% could be removed
 In this respect it is interesting to note that
 quantum Monte Carlo calculations of the lowest natural orbitals show that 
 they are very similar to the occupied Kohn-Sham orbitals \cite{Needs}. 
 Hence the corresponding natural Wannier functions are also expected to be 
 very similar to their Kohn-Sham counterparts. This similarity is presumably 
 no longer valid for the Wannier functions arising from the unoccupied bands. 
 Whereas the virtual Kohn-Sham Wannier functions are much less localized than 
 the occupied ones \cite{marzari_priv} this seems not to be the case for the 
 natural Wannier functions. 
%% could be removed

We now want to show under what conditions the natural Wannier functions
are exponentially localized. The key input to the corresponding proofs for
the ordinary Wannier functions \cite{Nenciu,desCloizeauxII} is the analyticity
of the projection operator onto the band states. The proof for energy-bands was
given in \cite{desCloizeauxI}. For the natural Wannier functions we have to 
prove the analyticity on a strip $K=\{{\bf k}'+i{\bf k}'',\; |{\bf k}''|<A\}$ 
of the projector $P({\bf k})$ onto an isolated set $\mathcal{B}$ of 
occupation bands. Because of the special properties of the density
matrix, the proof is straightforward. $D_{\bf k}$ is hermitian for real 
${\bf k}$; $N$-representability \cite{NatOrbt} requires for the eigenvalues 
$n_i\in [0,1]$; moreover, since a unit cell contains only a finite number of 
electrons and the bands are continuous, ${\rm Tr}\:D_{\bf k}$ is finite, 
i.e.\ $D_{\bf k}$ is trace class, and therefore, in particular, compact. 
If we assume that $D_{\bf k}$ is analytic on $K$, we can apply
the analytic Fredholm theorem \cite{ReedSimon}, which guarantees that
the resolvent of $D_{\bf k}$ is meromorphic on $K$ with the residues at the
eigenvalues being finite rank operators. Choosing a contour $C$ in $K$, which
exclusively encircles all occupation numbers in $\mathcal{B}$ (here we use that
the bands in $\mathcal{B}$ are isolated, i.e.\ they do not intersect with, or 
touch any band not in $\mathcal{B}$), we obtain the occupation band projector
\begin{equation}
 P({\bf k})={1\over2\pi i}\int_C {d\eta\over \eta-D_{\bf k}}\:,
\end{equation}
which, as desired, is analytic on $K$. Given the analytic band projection
operator, the proofs given in \cite{Nenciu,desCloizeauxII}, with the 
Hamiltonian replaced by the one-body density matrix, guarantee the
existence of exponentially localized Wannier functions if $\mathcal{B}$ contains
only a single band, or, for complex bands $\mathcal{B}$, in the tight-binding 
limit and in perturbation theory (of $D_{\bf k}$) around a situation, where
there do exist exponentially localized Wannier functions.

In the proof given above, we assumed that the ${\bf k}$-dependent
density matrix is analytic in ${\bf k}$. It is then natural to ask
what the analyticity of $D_{\bf k}$ means. Let us therefore distinguish
two cases: Either $D_{\bf k}$ is analytic in  ${\bf k}$
or it has a non-analyticity at some ${\bf k}={\bf k_F}$. Then, in the 
absence of degeneracies, in the first case, the occupation band structure 
$n_j({\bf k})$ will be analytic in ${\bf k}$ \cite{ReedSimon}, while in the
second case it will have a non-analyticity at ${\bf k_F}$. In the first case
there is no Fermi surface, and we thus associate this case with an insulator. 
As for the second case, assuming that perturbation theory holds, 
a discontinuity in $n_j({\bf k})$
implies a Fermi liquid, while an algebraic singularity would point to 
Luttinger liquid behavior. We thus associate this case with a metal. 
But we note that we are not aware of a proof that a discontinuity 
in $n_j({\bf k})$ by itself guarantees metallic behavior even when
perturbation theory breaks down \cite{Millis}. 
It is, however, hard to imagine an insulator with such a discontinuity, 
in particular, since it would also show up in the momentum distribution 
$N({\bf p}) = \int {\bf dr} {\bf dr}' 
\exp(-i {\bf p}({\bf r}-{\bf r}')) D({\bf r},{\bf r}')$, which is, in 
contrast to $n_j({\bf k})$, an experimentally accessible quantity.
From the Paley-Wiener theorem \cite{ReedSimon} (see also \cite{desCloizeauxI}),
it follows that for an insulator the 
density matrix, being the Fourier transform of $D_{\bf k}({\bf r},{\bf r}')$, 
decays exponentially with increasing distance $|{\bf r}-{\bf r}'|$. In the
three-dimensional case, the decay constant will in general be different along
different directions. For a metal, on the other hand, where 
$D_{\bf k}({\bf r}',{\bf r})$ is non-analytic on the Fermi surface, the
density matrix decays only algebraically. The decay properties of the
density matrix for interacting systems are thus qualitatively the same as
for non-interacting systems \cite{desCloizeauxI}. 
%% could be removed
 We emphasize that this result is valid for zero temperature, while at finite
 $T$ the decay should become exponential also for a metal \cite{finiteT}.
%% could be removed

\begin{figure}
 \centerline{\resizebox{3in}{!}{\includegraphics{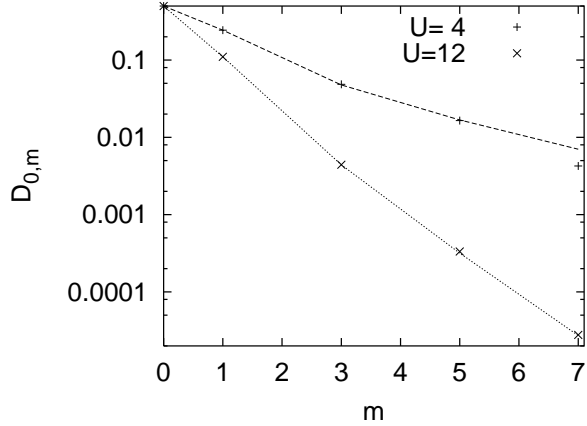}}}
 \caption[]{\label{decay}
            Exponential decay of the density matrix for the Hubbard chain.
            The symbols give the numerical values of the density matrix
            for a chain of 16 sites. The lines connect the values of 
            Ansatz (\ref{ansatz}), using the decay constants given in
            (\ref{gamma}).}
\end{figure}

As a specific and important example, we now analyze the localization 
properties of the density-matrix for a prototypical correlated system, 
the one-dimensional Hubbard model (with lattice constant $a$ and 
nearest-neighbor hopping matrix element $t$) \cite{Hubbard}
\begin{equation}
 H=-t\sum_{\langle i,j\rangle, \sigma}
     c^\dagger_{j\sigma}c^{\phantom{\dagger}}_{i\sigma} 
  + U\sum_i n_{i\uparrow} n_{i\downarrow} \: .
\end{equation}
For $U=0$ the system is metallic, while for any finite $U$ it is a Mott 
insulator \cite{LiebWu}.
Motivated by the result shown in Fig.\ \ref{decay} we choose as an Ansatz
the product of the exact density matrix for $U=0$ and 
an exponential factor:
\begin{equation}\label{ansatz}
  D_{0,m}=\langle\Psi|c^\dagger_{m\sigma} 
                      c^{\phantom{\dagger}}_{0\sigma}|\Psi\rangle
         ={\sin(\pi m/2)\over\pi m}\;e^{-\gamma a|m|} .
\end{equation}

\begin{figure}
 \centerline{\resizebox{3in}{!}{\includegraphics{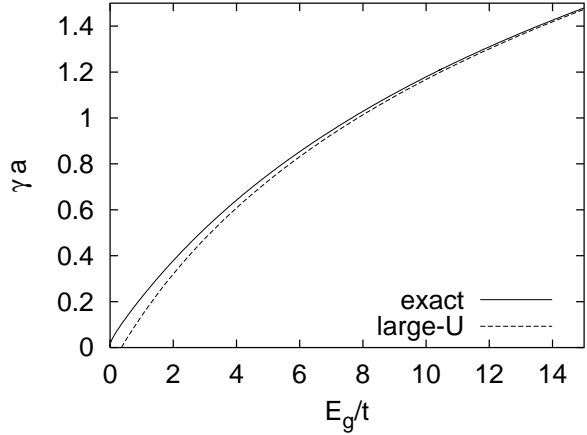}}}
 \caption{\label{ga}
          Decay constant $\gamma a$ of the density matrix for the Hubbard
          chain as function of the gap $E_g$. The dashed line shows the
          large-$U$ approximation (\ref{largeU}).} 
\end{figure}

To estimate the decay constant, we calculate $D_{0,1}$ from the exact 
ground-state energy of the Hubbard chain\cite{LiebWu}. Using the 
Hellmann-Feynman theorem we find 
\begin{equation}
 D_{0,1}=-{1\over4}\epsilon_{\rm kin}(U)/t
        =-{1\over4}\left(1-U{d\over dU}\right)\epsilon(U)/t ,
\end{equation}
which, together with the Ansatz (\ref{ansatz}), yields
\begin{equation}\label{gamma}
 \gamma a=
   -\ln\left(\pi\int_0^\infty dx\,{J_0^2(x)-J_1^2(x)\over1+e^{Ux/2t}}\right) , 
\end{equation}
where $J_0$ and $J_1$ are Bessel functions.
Fig.\ \ref{ga} shows a plot of $\gamma a$ as a function of the gap
$E_g/t=U/t-4+8\int_0^\infty dx J_1(x)/(x(1+\exp(Ux/2t)))$. 
For large $E_g$ we find the asymptotic behavior
\begin{equation}\label{largeU}
 \gamma a=-\ln\left({2\pi\ln(2)\over E_g/t+4}\right) \: ,
\end{equation}
while for small gap there exists no expansion, as for the Hubbard chain the
point $U=0$ is non-analytic.
The behavior of $\gamma a$ is thus qualitatively different in both the large 
and small gap limit from the analytical results for non-interacting systems
(cf.\ Fig.\ 1 in Ref.\ \cite{Arias}), although the overall shapes of the 
curves look similar. We note that the decay of the density matrix (\ref{ansatz})
has a power-law exponent different from the universal exponent for the 
non-interacting density-matrix found in \cite{HeVanderbilt}.

\begin{figure}
 \centerline{\resizebox{3in}{!}{\includegraphics{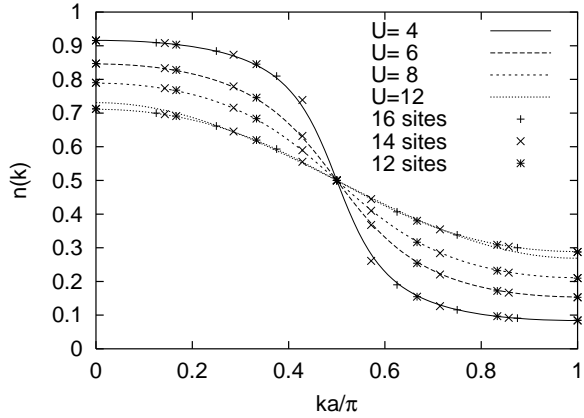}}}
 \caption{\label{nkplot}
          Occupation band structure for periodic Hubbard chains (determined by 
          exact diagonalization) compared to (\ref{nk}) with $\gamma a$ 
          obtained from (\ref{gamma}). All curves still differ
          significantly from the expression for $n(k)$ in the large-$U$ limit 
          \protect\cite{largeU}, which, for $U=12$, is plotted for comparison.}
\end{figure}

Fourier transforming the Ansatz (\ref{ansatz}), we obtain
\begin{equation}\label{nk}
 n(k)={1\over2}+{1\over\pi}\arctan\left({\cos(ka)\over\sinh(\gamma a)}\right) .
\end{equation} 
By construction, this is the step function for $U\to0$, while in the limit of 
large $U$ we have $n(k)={1\over2}+{4t\ln(2)\over U}\cos(ka)$; i.e.\ our Ansatz 
(\ref{ansatz}) is exact, both for $U\to0$ and $U\to\infty$ \cite{largeU}. 
For intermediate $U$, we compare (\ref{nk}) with the momentum distribution 
obtained from exact diagonalization of finite Hubbard chains.
This is shown in Fig.\ \ref{nkplot}: For $U$ large, but still far from the
large-$U$ limit, the agreement is perfect, and even for fairly small $U$, where
the decay length $1/\gamma a$ is of the order of the chain length, the
agreement between (\ref{nk}), which was derived for the infinite chain, and 
the results for the finite rings is amazingly good. 

The half-filled Hubbard chain is a Mott insulator for $U>0$, only for $U=0$
it is a metal. As expected, this is reflected in the decay properties of the
density matrix. While for $U=0$ it decays as $1/m$, for any finite $U$ the 
decay is exponential. It is interesting to compare the decay constant 
$\gamma a$ with the localization length $\lambda/a=\sqrt{d}/2\pi$, with 
$d=-\lim_{N\to\infty}N\ln|z_N|^2$ and 
$z_N=\langle\Psi_0|e^{{2\pi i\over N}\sum\hat{x}_i}|\Psi_0\rangle$, as defined
in \cite{Resta}. We find that both criteria describe
the metal-insulator transition correctly. 
As a surprising fact we note that $1/\lambda$ seems to be linear in $U$ 
down to very small values of $U$, with the constant of proportionality 
given by a large $U$ expansion.

Away from half-filling we have tried an Ansatz in the spirit
of (\ref{ansatz}) of the form ${\sin(k_F am)\over \pi m}\,|m+1|^{-\alpha}$.
Fourier transforming leads to an $n(k)$ with a Luttinger-like singularity with
exponent $\alpha$ at $k_F$. It fails, however, to also produce a singularity 
at $3k_F$ \cite{OgataShiba}.

To summarize, we have defined natural Wannier functions for interacting
systems from the eigenfunctions of the density matrix, and have shown 
under what conditions they can be proven to be exponentially localized.
The natural Wannier functions provide an optimal, localized basis for
describing a correlated system. In terms of computational efficiency, they
will allow the use of $O(n)$ methods in many-body calculations. We also 
expect them to provide understanding of the bonding in correlated solids.
In addition, we have shown that the characterization of metals and insulators 
by the decay properties of the density matrix does also apply to 
interacting systems, at least as long as perturbation theory holds.
Finally, we have investigated the one-dimensional Hubbard model,
proposed an expression for the density matrix of that model, and, for this
Mott insulator, derived a relation between the decay constant of the 
density matrix and the gap.

We thank L.N.~Trefethen for helping in identifying the relevant mathematical 
theorems, and T.~Arias, O.~Gunnarsson, P.~Horsch, J.~Hutter, R.M.~Martin, 
G.~Stollhoff, and D.~Vanderbilt for interesting discussions.

Note added in proof: After finishing the present work we became aware of
the paper \cite{kudinov}, in which also introduces natural Wannier functions.
We note, however, that in that paper the analyticity of the Bloch 
functions for simple bands is take for granted, rather than proven.


\begin{thebibliography}{00}
\bibitem{Kohn96}
W. Kohn, Phys. Rev. Lett. {\bf 76}, 3168 (1996);
W. Kohn, Phys. Rev. {\bf 133}, A171 (1964).
\bibitem{Arias}
S. Ismail-Beigi and T.A. Arias, Phys. Rev. Lett. {\bf 82}, 2127 (1999).
\bibitem{Baer}
R. Baer and M. Head-Gordon, Phys. Rev. Lett. {\bf 79}, 3962 (1997);
P. Maslen, C. Ochsenfeld, C. White, M. Lee, M. Head-Gordon, 
J. Chem. Phys. A {\bf 102}, 2215 (1998).
\bibitem{Marzari}
N. Marzari and D. Vanderbilt, Phys. Rev. B {\bf 56}, 12847 (1997).
\bibitem{GoedeckerRMP}
S. Goedecker, Rev. Mod. Phys. {\bf 71}, 1085 (1999).
\bibitem{Ordejon}
P. Ordejon, Comp. Mat. Sc. {\bf 12}, 157 (1998).
\bibitem{Resta}
R. Resta and S. Sorella, Phys. Rev. Lett. {\bf 82}, 370 (1999);
A.A. Aligia and G. Ortiz, Phys. Rev. Lett. {\bf 82}, 2560 (1999);
I. Souza, T. Wilkens, and R.M. Martin, Phys. Rev. B {\bf 62}, 1666 (2000).
\bibitem{Blount}
E.I. Blount, Solid State Physics {\bf 13}, 305 (1962).
\bibitem{NenciuRMP}
G. Nenciu, Rev. Mod. Phys. {\bf 63}, 91 (1991).
\bibitem{KingSmith}
R.D. King-Smith and D. Vanderbilt, Phys. Rev. B {\bf 47}, 1651 (1993);
                                   Phys. Rev. B {\bf 48}, 4442 (1993).
\bibitem{Nenciu}
G. Nenciu, Commun. Math. Phys. {\bf 91}, 81 (1983).
\bibitem{desCloizeauxII}
J. des Cloizeaux, Phys. Rev. {\bf 135}, A698 (1964).
\bibitem{Kohn59}
W. Kohn, Phys. Rev. {\bf 115}, 809 (1959).
\bibitem{desCloizeauxI}
J. des Cloizeaux, Phys. Rev. {\bf 135}, A685 (1964).
\bibitem{NatOrbt}
J. Coleman, Rev. Mod. Phys. {\bf 35}, 668 (1963);
E.R. Davidson, Adv. Quantum Chemistry {\bf 6}, 235 (1972).
\bibitem{AshcroftMermin}
N. Ashcoft and N.D. Mermin {\em Solid State Physics} 
(Saunders College, Philadelphia, 1976).
%% could be removed
 \bibitem{Needs}
 P. Kent, R. Hood, M. Towler, R. Needs and G. Rajagopal, 
 Phys. Rev. B {\bf 57}, 15293 (1998).
 \bibitem{marzari_priv}
 N. Marzari, private communication (1999)
%% could be removed
\bibitem{ReedSimon} 
M. Reed and B. Simon {\it Methods of Modern Mathematical Physics}
(Academic Press, New York, 1980);
T. Kato {\it Perturbation Theory for Linear Operators}
(Springer, New York, 1966). 
\bibitem{Millis}
A.J. Millis and S.N. Coppersmith, Phys. Rev. B {\bf 43}, R13770 (1991).
%% could be removed
 \bibitem{finiteT}
 S. Goedecker, Phys. Rev. B {\bf 58}, 3501 (1998).
%% could be removed
\bibitem{Hubbard}
M.C. Gutzwiller, Phys. Rev. Lett. {\bf 10}, 159 (1963);
J. Hubbard, Proc. Roy. Soc. London, Ser.\ A {\bf 276}, 238 (1963);
J. Kanamori, Prog. Theor. Phys. {\bf 30}, 275 (1963).
\bibitem{LiebWu}
E.H. Lieb and F.Y. Wu, Phys. Rev. Lett. {\bf 20}, 1445 (1968).
\bibitem{HeVanderbilt}
L. He and D. Vanderbilt, {\tt cond-mat/0102016}.
\bibitem{largeU}
J. Carmelo and D. Baeriswyl, Phys. Rev. B {\bf 37}, 7541 (1988).
\bibitem{OgataShiba}
M. Ogata and H. Shiba, Phys. Rev. B {\bf 41}, 2326 (1990).
\bibitem{kudinov}
E.K. Kudinov, Fiz. Tverd. Tela (St. Petersburg) {\bf 41}, 1582 (1999)
[Physics of the Solid State {\bf 41}, 1450 (1999)].
\end{thebibliography}
\end{document}